\documentclass[twocolumn,showpacs,preprintnumbers,superscriptaddress,amsmath,amssymb]{revtex4}
\usepackage{amsfonts}
\usepackage{amsmath}
\usepackage{amssymb}
\usepackage{graphicx}%
\usepackage{footmisc}
\usepackage{bm}
\usepackage{braket}
\usepackage{CJK}
\usepackage{hyperref}
\usepackage{cleveref}
\crefname{equation}{Eq.}{Eqs.}

\newcommand{\Tr}{\text{Tr}}

\newcommand{\Hc}{\text{H . c} }
\begin{document}
\begin{CJK*}{UTF8}{gbsn}
\title{Non-Markovian quantum dynamics: Extended correlated projection superoperators}

\author{Zhiqiang Huang (黄志强)}
\email{hzq@wipm.ac.cn}
\affiliation{Wuhan Institute of Physics and Mathematics, Chinese Academy of Sciences, Wuhan 430071, China}
\affiliation{University of the Chinese Academy of Sciences, Beijing 100049, China}

\date{\today}

\begin{abstract}
 The correlated projection superoperator techniques provide a better understanding about how correlations lead to strong  non-Markovian effects in open quantum systems.  Their superoperators are  independent of  initial state, which may  not be suitable for some cases. To improve this, we develop a new approach, that is extending the composite system before use the correlated projection superoperator techniques. Such approach allows choosing different  superoperators  for different initial states. We apply these techniques to a simple model to illustrate the general approach. The numerical simulations  of the full Schr\"{o}dinger equation of the model  reveal the power and efficiency of the method.
\end{abstract}


\maketitle
\end{CJK*}

\section{Introduction}
In both experimental and theoretical research, people are often interested in only  parts of objects, but there are always interactions between the parts of the whole.  Hence, the open system is common and important. A detailed review of open quantum systems can be found in \cite{BP02,VA17}. The dynamics of system usually satisfy the Markov approximation, which makes such processes history independent. In quantum dynamics, the evolution of such systems is described by the well-known Lindblad equation. As our ability to observe and control quantum system increases, non-Markovian behavior is becoming increasingly important. 

The  Nakajima-Zwanzig (NZ) equation and  time-convolutionless (TCL) equation are  powerful tools to analyze the open quantum systems.  With the Markov approximation, they can provide a Lindblad master equation.
 Without this approximation, the equations can be used to describe non-Markovian behavior.  Both of them are widely used in the research of open systems.   To solve the equations, one often needs  perturbation expansion with respect to the system-environment coupling. Approximations of the same order via these two methods can have very different physical meanings and ranges of applicability. In general, the TCL techniques are better \cite{R03}.

The NZ and TCL equations often  use the projection-operator method  to derive the dynamic equation of the system.  The original projection-operator method uses a projection superoperator (PS). The superoperator maps the total density matrix to a tensor product state, which is called the relevant part.  Total density matrix minus its relevant part  represents the irrelevant part.  The tensor product state omits any correlations  between system and environment. However, the correlations play a key role in non-Markovian behavior of open systems. This makes the PS method not  suitable for the research of non-Markovian behavior. 

 To study open quantum systems with  highly non-Markovian behavior, the correlated projection superoperator (CPS) techniques were  proposed \cite{BGM06,B07}. The CPS techniques  map the total density matrix to a quantum-classical state. This enables the relevant part to contain  classical correlations  and some quantum correlations, which may lead highly non-Markovian behavior.  However, the quantum discord and entanglement between system and environment is still lacking in the relevant part. A detailed review of correlations can be found in \cite{MPSVW10,S15}. Comparing with PS techniques, CPS techniques  generally provide a better understanding about the influence of the correlations between system and environment. It also yields accurate results already in the lowest order of the perturbation expansion \cite{BGM06,FB07}.  However, the accuracy of this approach  relies on the choice of  projection superoperators. An inappropriate  superoperator can lead  to disastrous results.

 The CPS techniques are very general.  To develop it, we first extend the environment space with an ancillary system. After that, we use CPS techniques in the enlarged space to get the evolution of relevant part. Although the relevant part here is still a quantum-classical state in the enlarged space, it allows the quantum discord between system and environment. Finally,  the evolution of  the system is obtained  by tracing over environment and ancillary system.  We call it extended correlated projection superoperator (ECPS) techniques. Compared with CPS, the ECPS provides more relevant variables. Hence, the relevant part permits more correlations. Besides that, the method may yield more accurate results.

 The basic idea underlying ECPS method is that one can separate the initial state into several pure states, and use different CPS to each pure state. This allows using the best CPS for each state; hence it probably yields more accurate results. If all the state share the same superoperator, the ECPS techniques will go back to the CPS techniques, which means the CPS techniques are good enough. The interaction of such systems often satisfies some conservation law. Such as conservation of energy \cite{BGM06}, or conservation of angular momentum \cite{FB07}. The conservation law often leads to classical correlations in those conserved quantities. If such conservation law is absent, as we will present in Sec. \ref{app},  then only the  ECPS techniques can ensure the accuracy of  the results.  Moreover, this approach allows the treatment of  initial states with non-vanished quantum discord by means of a homogeneous NZ or   TCL equation.

A homogeneous equation is easier. However, the form of  superoperator is mainly determined by interaction and steady state. A homogeneous master equation given by PS method is only applicable for some special initial states. For other states, even if the irrelevant part can be vanished under some special  superoperators. Their solution  may diverge in the limit $t\to\infty$.  In such cases, the homogeneous master equation given by the standard approach fails in any finite order of the
coupling strength. With ECPS method, the choice of  projection superoperators is more flexible. Therefore, it's easier to obtain a  convergence results under the homogeneous master equation. 

 We organize the paper as follows. In Sec. \ref{POT} we briefly review
 the standard projection-operator  method and the CPS techniques. We also review how to formulate the NZ equation and the TCL  equation. After that, we introduce the ECPS techniques and provide a  rough criterion to determine whether it needed. With the ECPS techniques, one can choose different projection superoperators for different initial states, which is totally different from the CPS method.
We also briefly discuss how to choose an appropriate  projection superoperator. In Sec. \ref{app} we apply the ECPS techniques to a  system-reservoir model. We discuss when the best projection superoperator is dependent on initial state.  Then,  we  show that any single projection superoperator can't ensure accurate results for all  initial states in some cases. This implies that the ECPS  techniques are necessary in these cases. 
After that, we explore the origin of the failure of  the homogeneous master equation. We show the ECPS techniques can provide a homogeneous master equation without divergence problems for more initial states. In Sec. \ref{dis} we conclude the paper and briefly discuss possible future developments of ECPS techniques.
\section{PROJECTION OPERATOR TECHNIQUES}\label{POT}
We consider an open quantum system $A$ is coupled with an environment $B$. Their Hilbert space of states are $\mathcal{H}_A$ and $\mathcal{H}_B$ respectively. The Hilbert space of states of the composite system is  $\mathcal{H}_A\otimes\mathcal{H}_B$. The dynamics of the total density matrix $\rho_{AB}$ of the composite system is governed by some Hamiltonian of the form $H=H_0+\alpha H_I$, where $H_0=H_A+H_B$  generates the free time evolution of the system and of the environment. $H_I$  describes the system-environment coupling. 
\subsection{The projection-operator method}
The PS techniques project the total density matrix of the composite system onto a tensor product state
\begin{equation}
  \mathcal{P}\rho_{AB}\equiv(\Tr_B\rho_{AB})\otimes \rho_B^0,
\end{equation}
where environment reference state $\rho_B^0$ is fixed in time.  Since $ \mathcal{P}^2=\mathcal{P}$, the superoperator  $\mathcal{P}$ is called projection superoperator.  The irrelevant part  is given by the complementary map $\mathcal{Q}=I-\mathcal{P}$, where $I$ denotes the identity map.  Obviously, the relevant part contains all the local information of the system $\Tr_B\mathcal{P}\rho_{AB}=\rho_A$. And being a tensor product state, the relevant part can't contain any system-environment correlation information.

The CPS techniques  project the total density matrix  onto a quantum-classical state
\begin{equation}\label{CPSsPO}
  \mathcal{P}\rho_{AB}\equiv\sum_\alpha \Tr_B(\Pi_{\alpha}^B\rho_{AB}) \otimes\rho_{\alpha}^B,
\end{equation}
where different $\alpha$ denotes different subspace $\mathcal{H}_\alpha$ of environment. They're orthogonal $\Pi_{\alpha}\Pi_{\beta}=\delta_{\alpha\beta}\Pi_{\beta}$
and complete $\sum_\alpha \Pi_{\alpha}=I^B$. $\Pi_{\alpha}^B$ is the identity matrix of $\mathcal{H}_\alpha$.   $\rho_{\alpha}^B$ is the reference state of $\mathcal{H}_\alpha$. If we take all the subspace $\mathcal{H}_\alpha$  as one dimension, then all reference state must be a projection operator $\rho_{\alpha}^B=\Pi_{\alpha}^B$. In such cases, the CPS becomes
\begin{equation}\label{CPSQC}
  \mathcal{P}\rho_{AB}\equiv\sum_i \Tr_B(\Pi_{i}^B\rho_{AB}) \otimes\Pi_{i}^B.
\end{equation}
The  collection of projection operators $\{\Pi_{j}^B|j\}$ determine the  projection superoperator, which directly affect the accuracy of the  lower order master equation. We shall  discuss it in detail in the  \cref{HCAPS}.  The definition of \cref{CPSsPO} and \cref{CPSQC} is  equivalent indeed: One can always  diagonalize $\rho_{\alpha}^B$ and redefine the basis of  $\mathcal{H}_\alpha$ to get \cref{CPSQC} from \cref{CPSsPO}. 

From the definition, $\Tr_B\mathcal{P}\rho_{AB}=\rho_A$ is  satisfied in CPS. Hence, its relevant part also contains all the local information of the system. Besides that, since its relevant part is quantum-classical state, it can contain some system-environment correlation information.

The relevant part provided by the CPS techniques contains more relevant variables, which make its dynamic equation more complex. Besides that, to determine the initial relevant part in CPS techniques, one may need more information about the environment.

\subsection{Nakajima-Zwanzig equation}
In the interaction picture with respect to $H_0$,  the von Neumann equation of the composite system can be written as
\begin{equation}\label{VNEIP}
  \frac{d}{dt}\rho_{AB}(t)=-i[\alpha H_I(t),\rho_{AB}(t)]\equiv\alpha\mathcal{L}(t)\rho_{AB}(t),
\end{equation}
where $H_I(t)=e^{iH_0t}H_Ie^{-iH_0t}$ is the Hamiltonian in the interaction picture and  $\mathcal{L}(t)$ denotes the corresponding Liouville superoperator. It's usually assumed that the
relations
\begin{equation}\label{oddv}
  \mathcal{P}\mathcal{L}(t_1)\mathcal{L}(t_2)\dots\mathcal{L}(t_{2n+1})\mathcal{P}=0
\end{equation}
hold for any  natural number $n$. Based on the von Neumann equation, one can derive an exact dynamical equation of relevant part \cite{BP02}
\begin{align} \label{NZ}
  \frac{d}{dt}\mathcal{P}\rho_{AB}(t)=\int_0^tdt_1\mathcal{K}(t,t_1)\mathcal{P}\rho_{AB}(t_1) \notag\\
  +\alpha\mathcal{P}\mathcal{L}(t)\mathcal{G}(t,0)\mathcal{Q}\rho_{AB}(0),
\end{align}
where superoperator
\begin{equation}
  \mathcal{K}(t,t_1)=\alpha^2\mathcal{P}\mathcal{L}(t)\mathcal{G}(t,t_1)\mathcal{Q}\mathcal{L}(t_1)\mathcal{P}
\end{equation}
is called the memory kernel or the self-energy.  $\mathcal{G}(t,t_1)=\mathcal{T}\exp[\alpha\int_0^t dt_2 \mathcal{Q}\mathcal{L}(t_2)]$   is  the propagator, where  $\mathcal{T}$ denotes chronological time ordering. \cref{NZ} is called the NZ equation. The inhomogeneous term $\mathcal{P}\mathcal{L}(t)\mathcal{G}(t,0)\mathcal{Q}\rho_{AB}(0)$ depends on the initial conditions at time $t=0$. It vanishes if the irrelevant part  vanishes at initial time. In PS techniques, if the initial state of the composite system is a tensor product state, one can choose the reference state $\rho_B^0=\rho_B(0)$ to make $\mathcal{Q}\rho_{AB}(0)=0$. In CPS techniques, if the initial state is quantum-classical state,  one can always let $\mathcal{Q}\rho_{AB}(0)=0$ by choosing an appropriate projection superoperator. From this perspective, a benefit of CPS method is that one can obtain a homogeneous master  equation for relevant part, even the initial state contains some  correlations between system and environment. 

Under the hypothesis (\ref{oddv}), when the relevant part at initial time vanishes, the lowest-order contribution is given by the second order
\begin{equation}
  \frac{d}{dt}\mathcal{P}\rho_{AB}(t)=\alpha^2\int_0^tdt_1\mathcal{P}\mathcal{L}(t)\mathcal{L}(t_1)\rho_{AB}(t_1).
\end{equation}

\subsection{Time-convolutionless master equation}
TCL master equation is an alternative way to deriving an exact master equation. It is a time-local equation of motion and doesn't depend on the full history of the system. The equation can be  written as
\begin{equation}\label{tcla}
  \frac{d}{dt}\mathcal{P}\rho_{AB}(t)=\mathcal{K}(t)\mathcal{P}\rho_{AB}(t)+\mathcal{I}(t)\mathcal{Q}\rho_{AB}(0).
\end{equation}
 $\mathcal{K}(t)$ is called the TCL generator, which can be expanded in terms of $H_I$.  The expansion can be  obtained from  ordered cumulants \cite{BP02}. By hypothesis  (\ref{oddv}), all  odd-order contributions vanish. The  second-order contribution reads
\begin{equation}\label{tclg2}
  \mathcal{K}_2(t)=\alpha^2\int_0^tdt_1\mathcal{P}\mathcal{L}(t)\mathcal{L}(t_1)\mathcal{P}.
\end{equation}
Combining \cref{tcla,tclg2}, if the inhomogeneous term vanishes, one  obtains 
\begin{equation}\label{tcl}
  \frac{d}{dt}\mathcal{P}\rho_{AB}(t)=\alpha^2\int_0^tdt_1\mathcal{P}\mathcal{L}(t)\mathcal{L}(t_1)\mathcal{P}\rho_{AB}(t).
\end{equation}
This is a second-order TCL master equation for relevant part.
The equations of motion provided by NZ and TCL techniques are different  in any finite order. But their exact solution should be the same.  Hence,  their accuracy is  of the same order.  In the following section, we will use TCL approach.

\subsection{Extended correlated projection superoperators}
The CPS techniques  use the most general linear projection superoperators that keeps all the local information of the system. However,  since it  uses a single superoperator for all the  initial states. One can not select the  projection superoperators  according to the  initial state.
Moreover,  its relevant part must lose some correlation, for instance, quantum discord and entanglement. Here we  propose a method that   allows choosing  superoperators for different initial states. One incidental  benefit is that the relevant part can contain quantum discord between system and environment now. The basic idea underlying our approach is the following. The initial  state of  the composite system  can always be separated into several  states $\rho^i_{AB}$. Its evolution comes directly from the evolution of these separated states $\rho_{AB}(t)=\sum_iP^i\rho^i_{AB}(t)$. If we use CPS separately, then choosing  superoperators  for different initial states is applicable. Moreover,  since using different  superoperators are permitted, the sum of those relevant part allows quantum discord, even the relevant part for each of them $\mathcal{P}^i\rho^i_{AB}$  doesn't contain quantum discord. 

The steps of the ECPS methods are as follows:
\begin{itemize}
  \item Extend the environment with an ancillary system and map the initial  state $\rho_{AB}=\sum_iP^i\rho^i_{AB}$ to a quantum-classical state $\rho_{ABC}=\sum_iP^i\rho^i_{AB}\otimes\Pi^i_{C}$.
  \item  Use CPS techniques in the extended space as
   $\mathcal{P}\rho_{ABC}=\sum_{i,j} \Tr_{BC}(\Pi_{i,j}^B\otimes\Pi_{i}^C\rho_{ABC}) \otimes\Pi_{i,j}^B\otimes\Pi_{i}^C$, where  the collection of projection operators $\{\Pi_{i,j}^B|j\}$ with different index $i$  provides a complete set  $\sum_j \Pi_{i,j}^B=I^B$. 
  \item Solve the master equation to get the evolution of relevant part $\mathcal{P}\rho_{ABC}(t)$.
  \item  The evolution of the system $\rho_A(t)$ can be obtained from the relevant part $\mathcal{P}\rho_{ABC}(t)$ by  partial tracing in the environment and ancillary system.
\end{itemize}
In this procedure,  the relevant part of each state is $\mathcal{P}^i\rho^i_{AB}= \Tr_{C}(\Pi_{i}^C\mathcal{P}\rho_{ABC})$. The ancillary space  denotes that different  states can use different CPS, i.e., $\{\Pi_{i,j}^B|j\}$ can be different for different $i$. Therefore, though $\mathcal{P}\rho_{ABC}(t)$ is still a quantum-classical sate,   $\Tr_C\mathcal{P}\rho_{ABC}(t)$ allows  quantum discord between system and environment.

One advantage of ECPS method is that its homogeneous master equation has a wider range of applications, such as cases that the initial state contains quantum discord between system and environment. 
A pure state is quantum correlated if and only if the state is entangled \cite{S15}. If the system isn't entangled with  environment initially, i.e., the initial state is separable, one can separate it into  several pure states. Since the composite system is isolated, those states remain  pure during the evolution. Hence, the pure state $\rho^i_{AB}$ won't contain quantum discord.  Correspondingly, the master equation of each pure state $\rho^i_{AB}$ can be homogeneous.  And the master equation of relevant part $\mathcal{P}\rho_{ABC}$ can still be homogeneous, even the initial state $\rho_{AB}(0)$ does contain quantum discord. 

Besides that, for pure state $\rho^i_{AB}$, its distance from separable states is the same as its distance from classical correlated states
\begin{equation}\label{DSEDC}
  \|\rho^i_{AB}-\mathcal{S}^i_{A:B}\|\equiv\min_{\sigma\in \mathcal{S}_{A:B}}\|\rho_{AB}^i-\sigma\|=\min_{\sigma\in \mathcal{C}_{A:B}}\|\rho_{AB}^i-\sigma\|,
\end{equation}
where $\mathcal{C}_{A:B}$ is the set of classical correlated states and  $\mathcal{S}_{A:B}$ is the set of separable states. According to \cref{DSEDC},  the irrelevant part in the ECPS techniques can be directly related to entanglement. This means that even the irrelevant part is non-trivial, the inhomogeneous term can always be upper bounded by the entanglement. An intriguing fact is that a general monogamy correlation measure beyond entanglement doesn't exist \cite{SAPB12}. Hence, it's impossible to find a similar result in the CPS techniques. 

The shortage of ECPS method is that the dynamical equation is more complicated. And  the entanglement between the system and the environment is still lacking in its relevant part. But, as we will explain below, include entanglement in relevant part may be unnecessary:
\begin{itemize}
  \item If the relevant part contains  all the local information of the system and all the system-environment correlation information, then it already contains  all system-related information in the composite system. Such relevant part is equivalent to the total density matrix of the composite system. It can be treated as a closed system and its evolution is unitary.
  \item The entanglement is monogamy. In many cases, the system-environment state is almost  indistinguishable from the separable state due to quantum de Finetti's theorem \cite{ZCZW15}.
  \item Though the relevant part of ECPS can't contain entanglement, one  can still take it into account with irrelevant part. The monogamy properties  of entanglement may be helpful to simplify the inhomogeneous term.
\end{itemize}
The PS techniques, CPS techniques and ECPS techniques project the total density matrix onto tensor product states, quantum-classical states, separable states respectively.  These approaches  together with  unitary evolution of closed systems compose a complete picture about the dynamics of the system, from the perspective of correlation. Each of them has its applications. The ECPS techniques can improve the accuracy of the lower-order equation, but also increase the complexity of equation. And It needs more information about the environment.

The dynamics of the open system is uniquely determined by the dynamical variables
\begin{equation}
  \rho_{i,j}(t)\equiv\Tr_{BC}(\Pi_{i,j}^B\otimes\Pi_{i}^C\rho_{ABC}(t)),
\end{equation}
from which the density matrix of the system reads
\begin{equation}
  \rho_s(t)=\sum_{i,j}\rho_{i,j}(t).
\end{equation}
The normalization condition is obviously satisfied
\begin{equation}
  \Tr_s\rho_s(t)=\sum_{i,j}\Tr_{ABC}(\Pi_{i,j}^B\otimes\Pi_{i}^C\rho_{ABC}(t))=1.
\end{equation}
Suppose  initial relevant part is vanished,  from NZ equation \cref{NZ},  the evolution equation of the dynamical variables becomes
\begin{equation}\label{tcl1}
  \frac{d}{dt}\rho_{i,j}(t)=\sum_k\int_0^tdt_1\mathcal{K}_{jk}^i(t,t_1)\rho_{i,k}(t_1),
\end{equation}
where the superoperator  
\begin{equation}
  \mathcal{K}_{jk}^i(t,t_1)\mathcal{O}_A\equiv \Tr_{BC}\{\Pi_{i,j}^B\otimes\Pi_{i}^C\mathcal{K}(t,t_1)(\mathcal{O}_A\otimes\Pi_{i,k}^B\otimes\Pi_{i}^C)\}.
\end{equation}
From TCL equation \cref{tcl},  we have
 \begin{equation}
  \frac{d}{dt}\rho_{i,j}(t)=\sum_k\mathcal{K}_{jk}^i(t)\rho_{i,k}(t),
 \end{equation}
 where the TCL generator is defined as
\begin{equation}
  \mathcal{K}_{jk}^i(t)\mathcal{O}_A\equiv \Tr_{BC}\{\Pi_{i,j}^B\otimes\Pi_{i}^C\mathcal{K}(t)(\mathcal{O}_A\otimes\Pi_{i,k}^B\otimes\Pi_{i}^C)\}.
\end{equation}
\cref{tcl1} can  be written as
\begin{equation}\label{tcl2}
  \frac{d}{dt}\mathcal{P}^i\rho_{i}(t)=\mathcal{K}^i(t)\mathcal{P}^i\rho_{i}(t),
\end{equation}
where $\mathcal{P}^i\rho_{AB}=\sum_j \Tr_{BC}(\Pi_{i,j}^B\rho_{AB}) \otimes\Pi_{i,j}^B$ and
\begin{equation}\label{DRI}
  \rho_{i}(t)=\Tr_{C}(\Pi_{i}^C\rho_{ABC}(t))=P^i\rho^i_{AB}.
\end{equation}
Comparing \cref{tcla} with \cref{tcl2}, it's easy to find that the ECPS method is just applying different  CPS  $\mathcal{P}^i$ to each pure state $\rho_{i}$. Similar to \cref{tcl},  the  second-order TCL equation in ECPS techniques can be written as
\begin{equation}\label{ECPSSO}
  \frac{d}{dt}\mathcal{P}^i\rho_{i}(t)=\alpha^2\int_0^tdt_1\mathcal{P}^i\mathcal{L}(t)\mathcal{L}(t_1)\mathcal{P}^i\rho_{i}(t).
\end{equation}

\subsection{Whether the ECPS techniques  are necessary}\label{WWETN}
A collection of projection operators can be parameterized as
\begin{equation}
  \Pi_{i,j}^{B }=U(\boldsymbol\theta^i)|j\rangle\langle j|U^\dagger(\boldsymbol\theta^i).
\end{equation}
The TCL equation \cref{tcla} is exact and its solution should be independent of the parameters $\boldsymbol\theta$. However,  the second-order TCL generator $\mathcal{K}^i_2(\boldsymbol\theta^i,t)$ is true depends on $\boldsymbol\theta$, so does its approximate solution. In principle,  the difference  $\mathcal{K}^i(t)-\mathcal{K}^i_2(\boldsymbol\theta^i,t)$ can tell  whether a projection superoperator is appropriate. However, it's hard to obtain $\mathcal{K}^i(t)$, which needs infinity order of expansion. 

An alternative way to judge a projection superoperator is from the unitary evolution of  the composite system
\begin{align} 
  \rho(t)=\mathcal{T}\exp[\alpha\int_0^t ds \mathcal{L}(s)]\rho(0)=(1+\alpha\int_0^t dt_1 \mathcal{L}(t_1)\notag\\
+\alpha^2\int_0^t dt_1 \int_0^{t_1} dt_2 \mathcal{L}(t_1) \mathcal{L}(t_2)+\dots)\rho(0).
\end{align}
Differentiating this equation with respect to time, we obtain
\begin{equation}\label{deocs}
  \partial_t\rho(t)=(\alpha\mathcal{L}(t)+\alpha^2\int_0^{t} dt_2 \mathcal{L}(t) \mathcal{L}(t_2)+\dots)\rho(0).
\end{equation}
The initial state of the composite system can be expressed as $\rho(0)=\mathcal{T}_\to\exp[-\alpha\int_0^t ds \mathcal{L}(s)]\rho(t)$, where $\mathcal{T}_\to$ denotes the antichronological time-ordering operator. 
Substituting it into \cref{deocs}, we obtain a time-local equation
\begin{equation}\label{tldeocs}
  \frac{d}{dt}\rho(t)=\mathcal{K}_{\text{tot}}(t)\rho(t),
\end{equation}
where superoperator $\mathcal{K}_{\text{tot}}$ is independent of projection superoperator. The error produced by expansion should be determined by the expansion order $\|\mathcal{K}^2_{\text{tot}}(t)-\mathcal{K}_{\text{tot}}(t)\|\sim O(\alpha^3)$. Results from the second order $\mathcal{K}^2_{\text{tot}}(t)$ should be pretty accurate if the perturbation theory is applicable in the composite system.  Hence, it may be appropriate to use the difference of superoperators
\begin{equation}\label{DOS}
  \mathcal{P}_{\theta}\mathcal{K}^2_{\text{tot}}(t)-\mathcal{K}_2(\boldsymbol\theta,t)
\end{equation}
to determine whether a projection superoperator is suitable. The distinguishability of superoperator can be  characterized by norm  \cite{K97}
\begin{equation}
  \|\Phi_0-\Phi_1\|_1\equiv\max\{\|\Phi_0(|\psi\rangle\langle\psi|)-\Phi_1(|\psi\rangle\langle\psi|)\|_1\},
\end{equation}
where $\||\psi\rangle\|=1$. A more precise norm \cite{K97} is 
\begin{equation}\label{NCPS}
  \|\Phi_0-\Phi_1\|_\lozenge\equiv\|(\Phi_0-\Phi_1)\otimes I\|_1.
\end{equation}
If the norm $\|\mathcal{P}_{\theta}\mathcal{K}^2_{\text{tot}}(t)-\mathcal{K}_2(\boldsymbol\theta,t)\mathcal{P}_{\theta}\|_\lozenge$ can be zero with parameters $\boldsymbol\theta$, then the collection of projection operators $\{\Pi^{B }(\boldsymbol\theta)\}$ is sufficiently accurate for all initial state. In such cases, the CPS techniques are good enough. Otherwise, the best projection superoperator may  depend on the initial state, and one may need ECPS techniques to improve the accuracy. 

The norm in \cref{NCPS} is based on the maximum distance of all states. It  determines whether the ECPS techniques are necessary, but can't tell which projection superoperator is the best for  a specific state.  The norm $\|(\mathcal{P}_{\theta}\mathcal{K}^2_{\text{tot}}(t)-\mathcal{K}_2(\boldsymbol\theta,t)\mathcal{P}_{\theta})\rho\|_1$  can reflect the accuracy for  a specific state. But, since the state of  the composite system can change during the evolution, such norm can't figure out which projection superoperator is better for the  whole process of evolution.  The exact relationship between the best projection superoperator and the initial states is beyond the scope of this paper. We leave this as an open question. 

Finding an exact relationship is difficult, but seeking candidates for some special cases  is easy. We will discuss this  in the following section.

\begin{figure}[htb]
  \centering
  \includegraphics[width=0.45\textwidth]{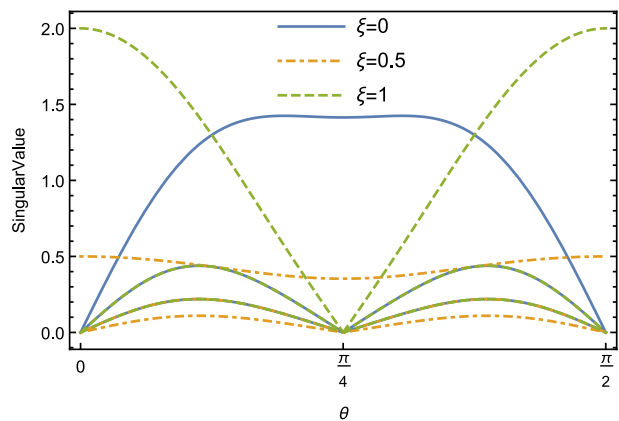}
  \caption{The singular values of Choi matrix  of  \cref{SDOS} under different projection superoperators  and  different interactions. For each $\xi$,  there are  at most three nontrivial singular values.  Parameters: $\lambda=\alpha^2\gamma N=1$.}
  \label{svwt}
  \end{figure}

\subsection{The  appropriate projection superoperator}\label{HCAPS}
In CPS techniques, one can construct the projection superoperators with the help of the conserved quantity of interaction $H_I$  \cite{FB07}. Suppose that $C$ is  a such conserved quantity, one can choose a collection of projection operators by the eigenstates of $C$
\begin{equation}
  \{\Pi_{i}^B:C|i\rangle=\lambda_i|i\rangle\}.
\end{equation}

Choosing an  appropriate projection superoperator for each state  $\rho^i_{AB}$ is also critical in ECPS techniques. Within the scope of ECPS method, the best  projection superoperator should depend on initial state. And there should be no conserved quantity as that used in the CPS techniques. If the interaction $H_I$ can be separated into several terms $H_I=\sum_i\alpha_iH_I^i $, and each term $H_I^i$ has  a conserved quantity  $C_i$, then one can construct the best  projection superoperator  with the eigenstates of one of the conserved quantities $C_k$ 
\begin{equation}
  \{\Pi_{i,j}^B:C_k|i,j\rangle=\lambda_j^k|i,j\rangle\}.
\end{equation}
Which $C_k$ to use,  from the interaction and initial state of the model, is still a problem. In the next section,  we will briefly discuss this issue with a  system-reservoir model. A systematic analysis is beyond the scope of this paper.

\section{Application}\label{app}
\subsection{The model}\label{TM}
To illustrate the general considerations of the previous sections, we apply the ECPS techniques to a  two-state system $S$. The model  was adapted from \cite{BGM06}. The two states of the system are degenerate. The system is coupled to an environment $E$, which consists of $N$ equidistant energy levels in an energy bands of width $\delta\epsilon$,  and each energy level is doubly degenerate. The total Hamiltonian of the model in the  Schr\"{o}dinger picture is $H=H_0+\alpha(V_1+V_2)$, where
\begin{equation}
  H_0=\sum_{n=1}^N\sum_{i=1,2}\frac{\delta\epsilon}{N}n|n,i\rangle\langle n,i|,
\end{equation}
and 
\begin{align}
  V_1=(1-\xi)\sum_{n_1,n_2}c(n_1,n_2)\sigma_+|n_1,1\rangle\langle n_2,2|+\text{H. c}, \notag\\
V_2=\xi\sum_{n_1,n_2}c'(n_1,n_2)\sigma_{\to}|n_1,+\rangle\langle n_2,-|+\text{H. c}.
\end{align}
Here $\sigma_+=(\sigma_x-i\sigma_y)/2$ and $\sigma_\to=(\sigma_y-i\sigma_z)/2$. $\{\sigma_i|i=x,y,z\}$ is the  standard Pauli matrices. The environment state $|n_1,\pm\rangle\equiv(|n_1,1\rangle\pm|n_1,2\rangle)/\sqrt{2}$. The coupling constant $c(n_1,n_2)$ and $c'(n_1,n_2)$ are independent and identically distributed complex Gaussian random variables satisfying 
\begin{equation}\label{MVOGRV}
  \begin{aligned}
  \langle c(n_1,n_2)\rangle=\langle c'(n_1,n_2)\rangle=0, \\
\langle c(n_1,n_2) c(n_3,n_4)\rangle=\langle c'(n_1,n_2)c'(n_3,n_4)\rangle =0, \\
\langle c(n_1,n_2) c^*(n_3,n_4)\rangle=\langle c'(n_1,n_2)c'^*(n_3,n_4)\rangle  \\
=\delta_{n_1,n_3}\delta_{n_2,n_4}, \\
\langle c(n_1,n_2) c'(n_3,n_4)\rangle=\langle c(n_1,n_2) c'^*(n_3,n_4)\rangle=0.
\end{aligned}
\end{equation}
It's easy to check that interaction operator $V_1$ and $V_2$ don't commute with each other  $[V_1,V_2]\neq0$. But their commutator is vanished in average of those random  variable, i.e.,  $\langle[V_1,V_2]\rangle=0$.

In  the interaction picture, the Hamiltonian of  von Neumann equation (\ref{VNEIP}) is $H_I(t)=V_1(t)+V_2(t)$, where 
\begin{align}
  V_1(t)=(1-\xi)(\sigma_+B(t)+\sigma_-B^\dagger(t));  \notag \\
 V_2(t)=\xi(\sigma_\to B'(t)+\sigma_\gets B'^\dagger(t))
\end{align}
and 
\begin{align}
  B(t)=\sum_{n_1,n_2}c(n_1,n_2)e^{-i\omega(n_1,n_2)t}|n_1,1\rangle\langle n_2,2|, \notag \\
  B'(t)=\sum_{n_1,n_2}c'(n_1,n_2)e^{-i\omega(n_1,n_2)t}|n_1,+\rangle\langle n_2,-|.
\end{align}
The energy difference $\omega(n_1,n_2)=\delta\epsilon(n_2-n_1)/N$. 

When $\xi=0$, the CPS techniques is enough, just as its predecessor. Suppose an environment operator $\mathcal{O}$ satisfies 
\begin{equation}
  \mathcal{O}|n_1,1\rangle=-|n_1,1\rangle;\mathcal{O}|n_1,2\rangle=|n_1,2\rangle.
\end{equation}
Then  $ \mathcal{O}+\sigma_z$ is a conserved quantity of Hamiltonian $H$. Following the principle mentioned in \cref{HCAPS}, the best projection superoperator can be constructed with $\Pi^{1,2}$, where $\Pi^i=\sum_{n}|n,i\rangle\langle n,i|$. Such superoperator is  the same as  that used in \cite{BGM06}. Similar, when $\xi=1$, the best projection superoperator  should be constructed with $\Pi^{+,-}$.  When $0<\xi<1$, there is no such conserved quantity anymore. As explained in \cref{HCAPS},  we need ECPS techniques in such cases. The best projection  superoperator should  be constructed with  $\Pi^{1,2}$ or $\Pi^{+,-}$. It depends on the specific initial state to decide which one is better.
\subsection{TCL master equation from ECPS techniques}
Suppose the projection superoperator in \cref{tcl2} is  
\begin{equation}
  \mathcal{P}_\theta\rho_{AB}=\sum_{i=1,2`} \Tr_{B}(\Pi^{i}_\theta\rho_{AB}) \otimes\Pi^{i}_\theta/N,
\end{equation}
where $\Pi^i _\theta=\sum_{n}|n,i,\theta\rangle\langle n,i,\theta|$ and 
\begin{equation}
  \left(
\begin{array}{c}
 |n,1,\theta \rangle  \\
 |n,2,\theta \rangle  \\
\end{array}
\right)= \left(
\begin{array}{cc}
 \cos \theta  &  \sin \theta \\
 \sin \theta  &  \cos \theta  \\
\end{array}
\right)\left(
  \begin{array}{c}
   |n,1\rangle  \\
   |n,2\rangle  \\
  \end{array}
  \right).
\end{equation}
With $\theta=0$, $\Pi^i _\theta$ reads $\Pi^{1,2}$. With $\theta=\pi/4$, it gives $\Pi^{+,-}$.  According to \cref{ECPSSO},  the second-order TCL generator can be expressed as
\begin{equation}\label{STCLG}
  \mathcal{K}^i(t)=\alpha^2\int_0^tdt_1\mathcal{P}_\theta  \langle\mathcal{L}(t)\mathcal{L}(t_1) \rangle\mathcal{P}_\theta.
\end{equation}
According to \cref{MVOGRV}, the random variables are independent of each other, so the total  generator can be obtained by just adding two generators of corresponding interaction 
\begin{equation}\label{TGAGT}
  \mathcal{K}^i(t)=\mathcal{K}^i_1(t)+\mathcal{K}^i_2(t),
\end{equation}
where  $\mathcal{K}^i_{1,2}(t)$ are the second-order TCL generator of $V_{1,2}$ respectively. Based on calculations in \cref{STG}, all terms in the generator $\mathcal{K}^i_{1,2}(t)$ share the same  time-dependent part $h(\tau)$, which may be approximated by a $\delta$ function when $\delta\epsilon\tau\gg1$. In this way, we obtain the following second-order TCL master equation 
\begin{equation}\label{SOTCLEM}
 \frac{d}{dt}\mathcal{P}_\theta\rho_{i}(t)
 =\alpha^2\gamma \mathcal{P}_\theta(\langle\mathcal{L}_1\mathcal{L}_1\rangle+\langle\mathcal{L}_2\mathcal{L}_2\rangle)\mathcal{P}_\theta\rho_{i}(t),
\end{equation}
where $\mathcal{L}_{1,2}$ are Liouville  superoperators of $V_{1,2}$ respectively. In the following, we write the elements of $\rho_{i}(t)$ as 
\begin{equation}
  \rho_{i}^{jl,km}(t)=\Tr(\rho_{i}(t)|km\rangle\langle jl|), \quad l,m=0,1;j,k=1,2,
\end{equation}
where $|km\rangle\langle jl|=|m\rangle_A\langle l|\otimes (\sum_{n}|n,k\rangle_B\langle n,j|)$.  By taking the partial trace over the environment, we obtain $\rho_{i}^{lm}(t)=\sum_{j}\rho_{i}^{jl,jm}(t)$. From the definition (\ref{DRI}), the  elements of the reduced density matrix satisfies $\rho_S^{lm}(t)=\sum_{i}\rho_{i}^{lm}(t)$.

If use $\mathcal{P}_{\theta=0}$, then according to Eqs. (\ref{P01DE}), the coherences $\rho_{i}^{01}(t)$ of the solution should decay  exponentially  over time.  If set  $\xi=0$ further, then populations  $\rho_{i}^{11}(t)$ of the solution  will depend on initial state. This indicates the strongly non-Markovian effect in this model. Moreover, the populations of steady state $\lim_{t\to\infty}\rho_{i}^{11}(t)$  also depend on initial state.

If use $\mathcal{P}_{\theta=\pi/4}$, then according to  Eqs. (\ref{P+-DE}), the populations of the solution should decay exponentially  to $I/2$ over time. If set   $\xi=1$ further, then the coherences of steady state can be nontrivial and depend on initial state.
 
When $\xi\neq0$, the  populations and coherences of the solution are both dependent on initial state, not matter which  superoperator one uses. In such cases, the dynamics of the system does not even  represent a semigroup. 
\subsection{The best projection superoperator may be depend on initial state}\label{BPSISD}
As discussed in \cref{WWETN}, one can use a generator $\mathcal{K}_{\text{tot}}$ to judge whether the ECPS techniques improve or not. In the following text, we will illustrate this with  the  model in \cref{TM}.

According to \cref{MVOGRV},  mean value of Liouville superoperator  vanishes: $\langle \mathcal{L}(t)\rangle=0$. Hence, \cref{deocs} becomes 
\begin{equation}
  \partial_t\rho(t)=(\alpha^2\int_0^{t} dt_2\langle  \mathcal{L}(t) \mathcal{L}(t_2)\rangle +O(\alpha^3))\rho(0).
\end{equation}
The lowest order of the generator $\mathcal{K}_{\text{tot}}$ is
\begin{equation}
  \mathcal{K}^2_{\text{tot}}(t)=\alpha^2\int_0^tdt_1 \langle\mathcal{L}(t)\mathcal{L}(t_1)\rangle.
\end{equation}
 Similar to \cref{TGAGT}, we have 
\begin{equation}\label{TGAGTT}
  \mathcal{K}^2_{\text{tot}}(t)\rho=\alpha^2\int_0^{t} dt_2 (\langle\mathcal{L}_1(t) \mathcal{L}_1(t_2)\rangle+\langle\mathcal{L}_2(t) \mathcal{L}_2(t_2)\rangle)\rho.
\end{equation}
When $\delta\epsilon\tau\gg1$, according to the  calculations in \cref{STG},  we have $\mathcal{P}_\theta\mathcal{K}^2_{\text{tot}}(t)\sim\mathcal{P}_\theta\mathcal{K}^2_{\text{tot}}$ and
\begin{equation}\label{STTG}
  \mathcal{K}^2_{\text{tot}}\rho=\alpha^2\gamma(\langle\mathcal{L}_1\mathcal{L}_1\rangle+\langle\mathcal{L}_2\mathcal{L}_2\rangle)\rho.
\end{equation}

With \cref{SOTCLEM,STTG}, the difference (\ref{DOS}) is  independent of time
\begin{equation}\label{SDOS}
  \Delta= \mathcal{P}_\theta\mathcal{K}^2_{\text{tot}}-\mathcal{K}_2(\theta).
\end{equation}
The singular values of the Choi matrix  of $\Delta$ are shown in  \cref{svwt}. The norm (\ref{NCPS}) vanishes if all the  singular values vanish. When $\xi=0$ or $\xi=1$, there always exists a $\theta$ to make all the singular values  zero. This means  that  the  projection superoperator $\mathcal{P}_\theta$ is suitable for all states. In such cases, the interaction  contains a conserved quantity, and the CPS techniques are good enough. If there is no such quantity, such as cases that $\xi=1/2$. Whatever $\theta$ one chooses, the  singular values  can not all be zero.  Under the circumstances, in any fixed projection superoperators,  the master equation is inaccurate for some  initial states. The best projection superoperator  will depend on  the initial state and  the ECPS techniques can yield  better results. As presented in \cref{svwt}, the projection superoperator with $\theta=0$ or $\theta=\pi/4$  leads to fewer and smaller nontrivial singular values. Hence, the best projection superoperator should be selected from $\mathcal{P}_{\theta=0}$ or  $\mathcal{P}_{\theta=\pi/4}$, which accords with the discussion in  \cref{TM}. 

  \subsection{Comparing the results}
  \begin{figure}[htb]
    \centering
    \includegraphics[width=0.45\textwidth]{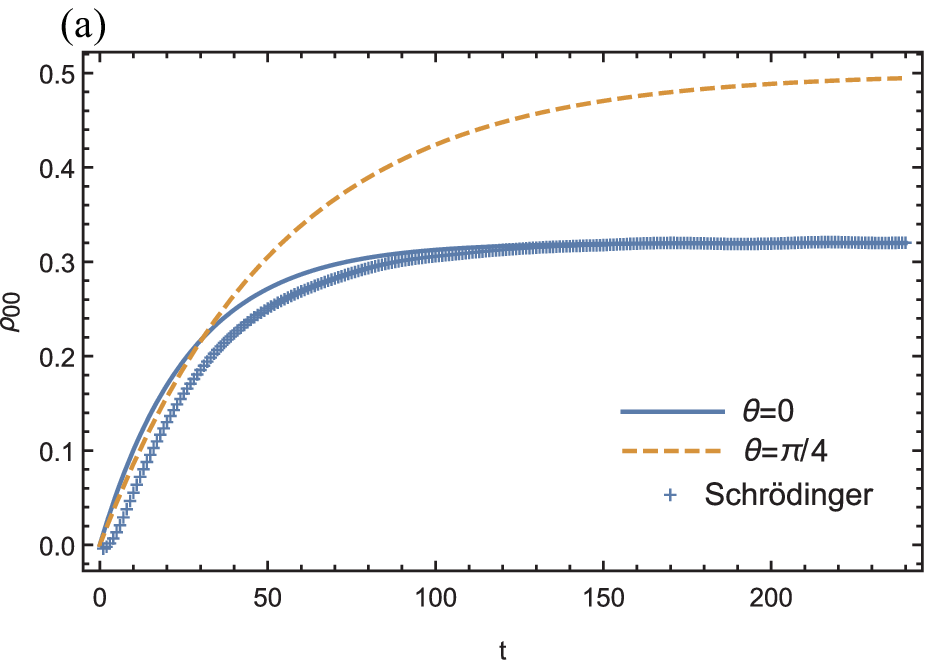}
    \includegraphics[width=0.45\textwidth]{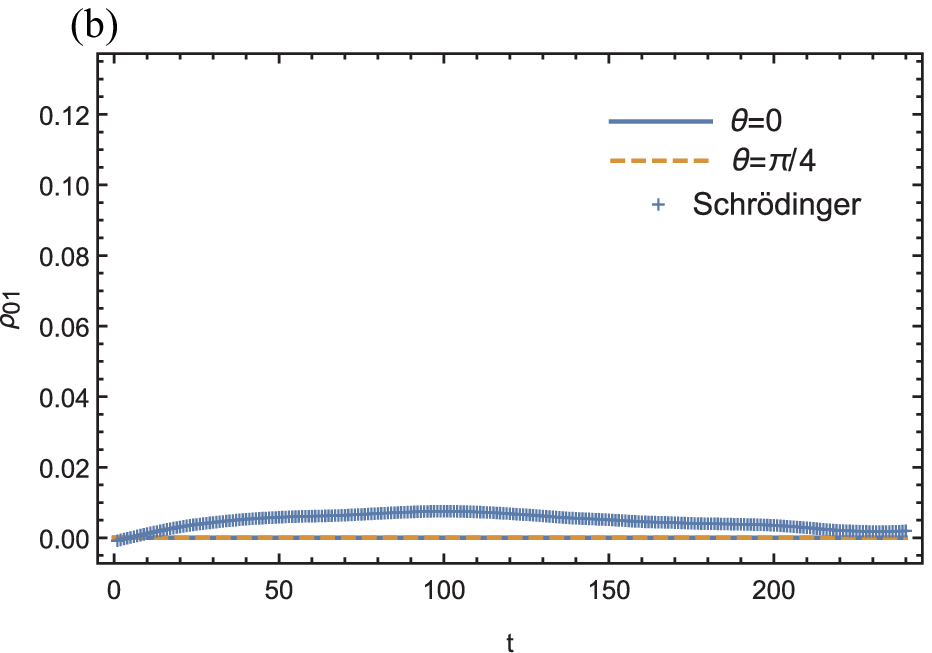}
    \caption{Comparison of the second-order TCL approximation using  different projection superoperators and of the numerical solution of the Schr\"{o}dinger equation. (a) and (b) show the evolution of  population $\rho_{00}$ and coherence $\rho_{01}$, respectively. 
    Here $\xi=0$ and the initial state is $|0\rangle_A\langle0|\otimes\Pi^1_\theta/N$, where $\sin(\theta)=3/5$. Other parameters: $N=60$, $\delta \epsilon =0.5$ and $\alpha=5\times10^{-3}$.}
    \label{e0}
    \end{figure}

    \begin{figure}[htb]
      \centering
      \includegraphics[width=0.45\textwidth]{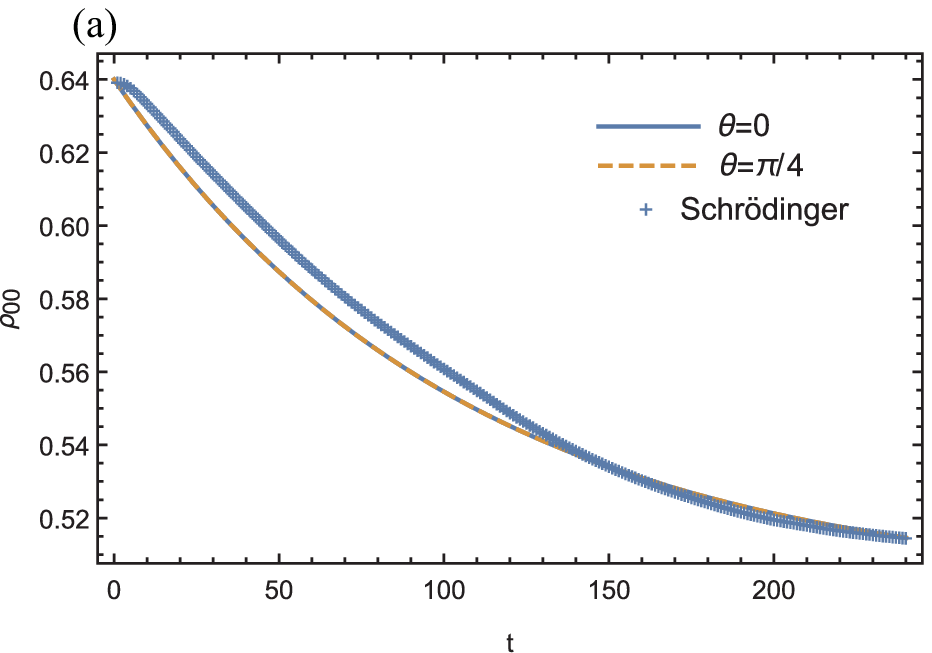}
      \includegraphics[width=0.45\textwidth]{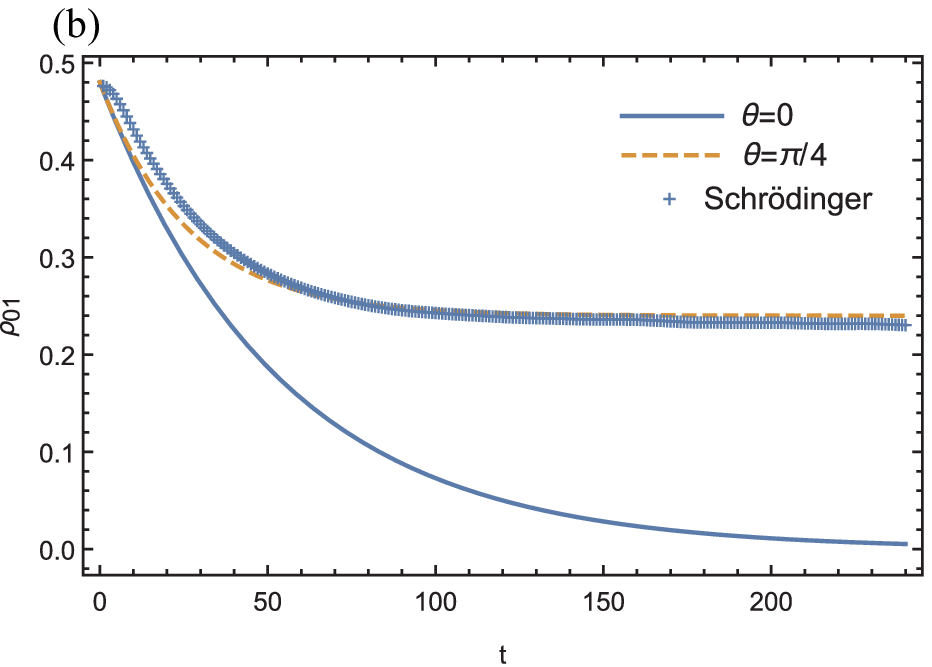}
      \caption{The same as \cref{e0} but $\xi=1$ and the initial state is $|\psi\rangle_A\langle\psi|\otimes\Pi^1$, where $|\psi\rangle=0.6|0\rangle+0.8|1\rangle$. }
      \label{e1}
      \end{figure}

      \begin{figure}[htb]
        \centering
        \includegraphics[width=0.45\textwidth]{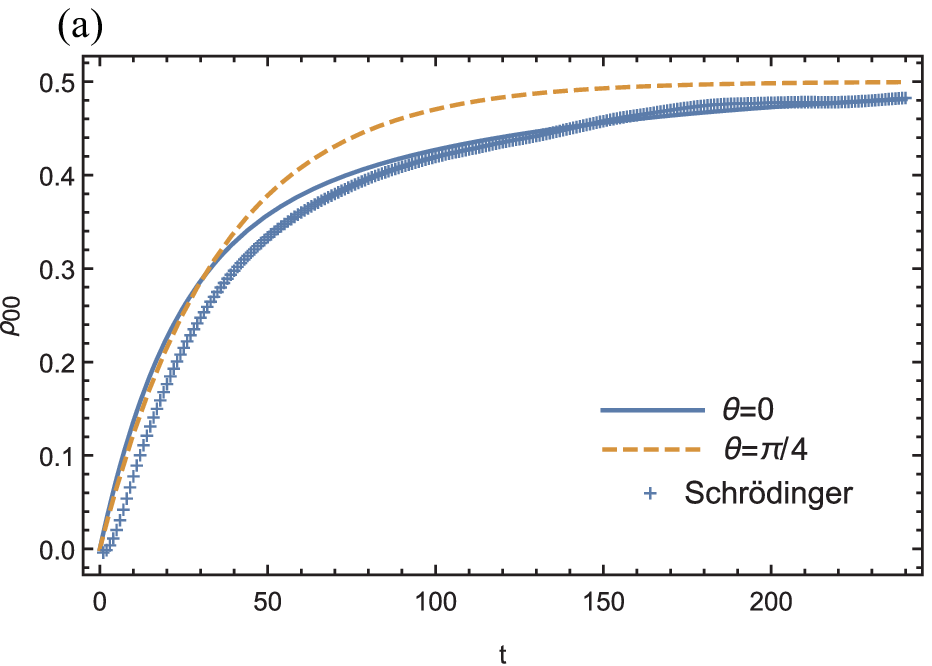}
        \includegraphics[width=0.45\textwidth]{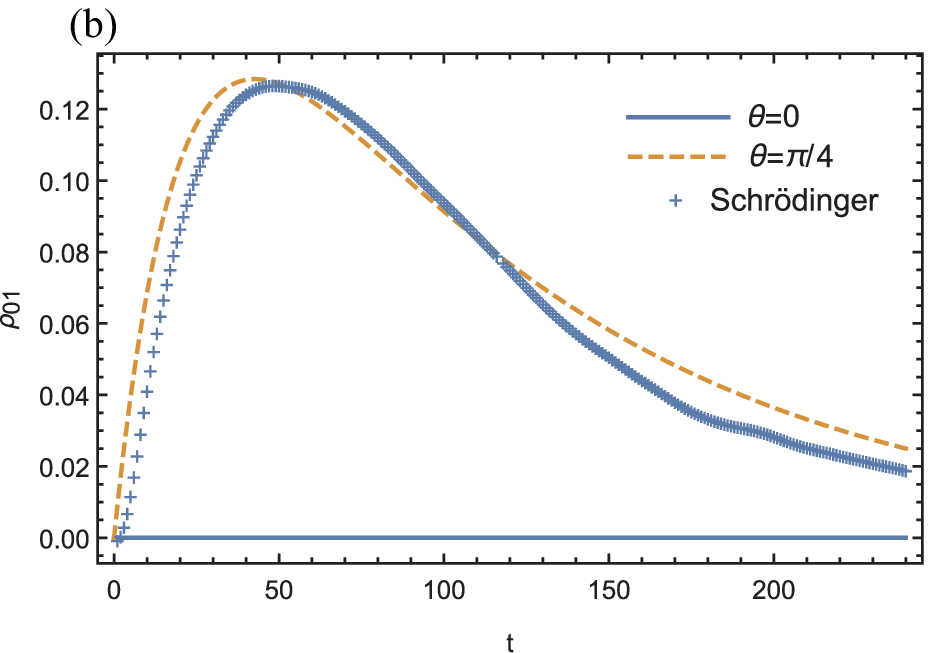}
        \caption{The same as \cref{e0} but $\xi=0.5$ and  $\alpha=10^{-2}$.}
        \label{e051}
        \end{figure}

        \begin{figure}[htb]
          \centering
          \includegraphics[width=0.45\textwidth]{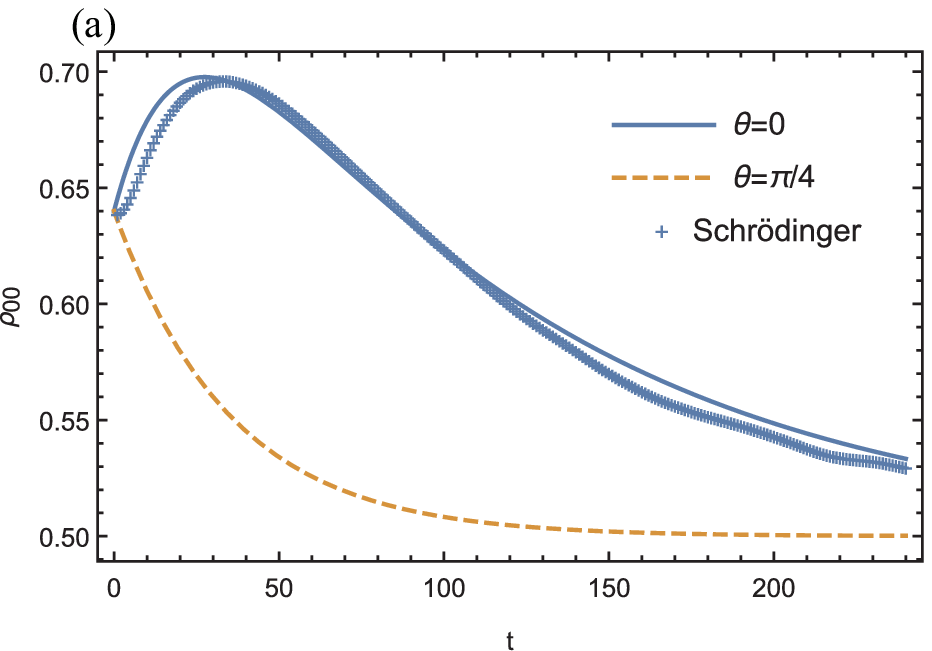}
          \includegraphics[width=0.45\textwidth]{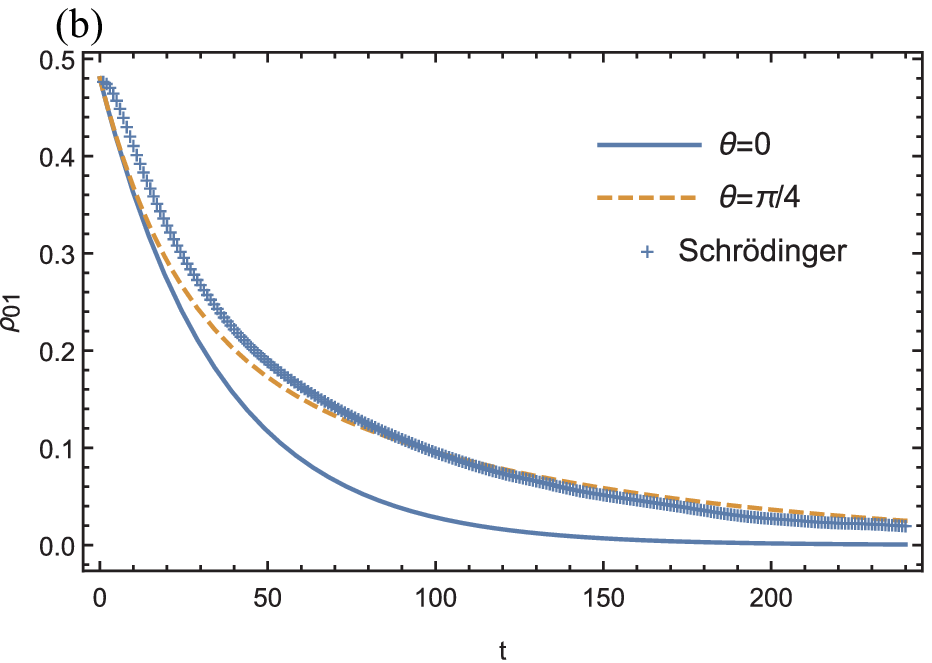}
          \caption{The same as \cref{e1} but $\xi=0.5$ and $\alpha=10^{-2}$.}
          \label{e050}
          \end{figure}

In this section, we compare the results obtained from different projection superoperators [ see \cref{TDE} for details] and  the numerical solution of the Schr\"{o}dinger equation.  
 
As shown in \cref{e0}, when $\xi=0$,  two  superoperators give the same coherences, which are consistent with the numerical solution. Their populations are totally different. The populations provided by $\Pi^{1,2}$ are pretty accurate  through  the whole process.  In contrast, the projection superoperator  $\mathcal{P}_{\theta=\pi/4}$ can't even provide an accurate  steady state. These results are consistent with \cref{svwt} and the discussions in  \cref{TM}: When $\xi=0$, the  projection superoperator  $\mathcal{P}_{\theta=0}$ is the best for all initial states.

As shown in \cref{e1}, when $\xi=0$,  the populations are the same and  consistent with the numerical solution. But the coherences are totally different. The coherences provided by $\mathcal{P}_{\theta=\pi/4}$ are pretty accurate.   And the projection superoperator $\mathcal{P}_{\theta=0}$ cannot  give an  accurate  steady state. These results are also consistent with \cref{svwt} and the  discussions in  \cref{TM}: When $\xi=1$, the  projection superoperator  $\mathcal{P}_{\theta=\pi/4}$ is the best for all initial states.

As shown in \cref{e051}, the  populations provided by two projection superoperators are very close. Both give a good approximation. Their  coherences are totally different. The coherences given by $\Pi^{+,-}$  are pretty accurate.   In contrast, the projection superoperator with  $\Pi^{1,2}$ provide a worse approximation. 
 In \cref{e050},  the  populations provided by two projection superoperators are very close. Both  give a good approximation. Their   coherences are totally different. The populations provided by $\Pi^{1,2}$ are pretty accurate, which can not be achieved by the  superoperator $\mathcal{P}_{\theta=\pi/4}$. 
 These results are consistent with \cref{svwt} and the  discussions in  \cref{TM} also: When $\xi=0.5$, the best projection superoperator should depend on initial state and can be selected from $\mathcal{P}_{\theta=0}$ or $\mathcal{P}_{\theta=\pi/4}$.

 Comparing \cref{e0} with \cref{e051} or \cref{e1} with \cref{e050}, one may conclude that a different interaction can completely change the best projection superoperator. 

\subsection{The failure of homogeneous master equation}
The Born-Markov approximation may fail even when the standard Markov condition is fulfilled \cite{BGM06}. In their model, the homogeneous master equation obtained by the PS techniques   is  inaccurate in the low-order expansion. And  the populations obtained in the high-order expansion may diverge in the limit $t\to\infty$. In this procedure, a higher order expansion can only improve the approximation for short term, but leads to nonphysical results for longer term. It was proven that the CPS techniques can perfectly solve the problem. 

We believe that the root  of the  problem is that  the  steady state given by  lowest order master equation is totally wrong.  If the Markov condition is fulfilled and there is always a steady state, the populations  obtained in  the lowest order master equation can be generally written  as 
\begin{eqnarray}
  \rho_i(t)\equiv\langle i|\rho(t)|i\rangle=A_ie^{-\gamma_1^it}+\rho^s_i,
\end{eqnarray}
where $\rho^s_i$ is the populations of steady state and all  the $\gamma_1^i$ is  positive.  In the high-order expansion, the evolution of populations  becomes
\begin{eqnarray}
  \rho_i(t)=A_ie^{-\gamma_1^it-\gamma_2^it^2+O(t^{3})}+\rho^s_i.
\end{eqnarray}
Since the master equation of higher order  should yield a better approximation for short term,  some of  $\gamma_2^i$ must be negative if  $\rho^s_i$ is wrong. However, the negative $\gamma_2^i$ can lead to divergence  for longer term. Hence, a wrong steady state may be the root of the problem. In Ref. \cite{BGM06}, the  homogeneous master equation obtained by the CPS techniques can yield an accurate steady state, which solves the divergence problem in higher order expansion.

Similar, the homogeneous master equation obtained by the CPS techniques may also fail in some cases. And some of them can be solved by ECPS techniques.  For instance,  the model of \cref{TM} with $\xi=0$. Supposing the initial state is  
\begin{equation}
  \rho_0=P_1\rho_A^1\otimes \frac{1}{2N}I+(1-P_1)\rho_A^2\otimes\frac{1}{N}\Pi^+.
\end{equation}
Since $I=\Pi^{1}+\Pi^{2}=\Pi^{+}+\Pi^{-}$,   in CPS techniques, one needs  projection superoperator  $\mathcal{P}_{\theta=\pi/4}$ to obtain a homogeneous master equation. In this approach,  the populations of the steady state  should be  $I/2$ according to  \cref{P+-DE}. While in ECPS techniques, one can choose $\mathcal{P}_{\theta=0}$ for the term $P_1\rho_A^1\otimes \frac{1}{2N}I$ and  $\mathcal{P}_{\theta=\pi/4}$ for the term $(1-P_1)\rho_A^2\otimes\frac{1}{N}\Pi^+$,  to get a  homogeneous master equation. Suppose 
\begin{equation}
  \rho_A^1=\left(
\begin{array}{cc}
 P & 0 \\
 0 & 1-P \\
\end{array}
\right).
\end{equation}
According to  \cref{P01DE}, the steady state of the term $P_1\rho_A^1\otimes \frac{1}{2N}I$ should be
\begin{eqnarray}
  \rho_{A,s}^1=\frac{1}{4}\left(
\begin{array}{cc}
 1+2P & 0 \\
 0 & 3-2P \\
\end{array}
\right).
\end{eqnarray}
According to the analysis of \cref{BPSISD}, when $\xi=0$, only the projection superoperator $\mathcal{P}_{\theta=0}$ can always give a good approximation. Hence, the  steady state given by CPS techniques would be wrong whenever  $P\neq1/2$. Since the steady state is wrong, the homogeneous master equation provided by the CPS techniques should fail. 

The projection superoperator  $\mathcal{P}_{\theta=\pi/4}$ can provide a good approximation for some states, such as \footnote{When $\xi=0$, though only $\mathcal{P}_{\theta=0}$ can give a good approximation for all initial states. $\mathcal{P}_{\theta=\pi/4}$  can still give a good approximation for some states.}
\begin{equation}
  \rho_A^2=\frac{1}{2}\left(
\begin{array}{cc}
 1 & A \\
 A^\dagger & 1 \\
\end{array}
\right).
\end{equation}
In these cases, since the ECPS techniques enable us to choose different projection superoperators for the separated states, it  solves the divergence problem in the CPS techniques. But, if $\mathcal{P}_{\theta=\pi/4}$  provides a false steady state to term $(1-P_1)\rho_A^2\otimes\frac{1}{N}\Pi^+$, then the ECPS techniques can also fail. 


In general, the homogeneous master equation from the  ECPS techniques is applicable  to more general  initial state.  The results are likely to be  convergence. Moreover, it yields  more accurate results  in lower order equation.

\section{Conclusion and outlook}\label{dis}
We find that the best projection superoperator can be dependent on initial state. In such circumstances, the ECPS techniques  yield a better approximation. Besides that, the relevant part in the ECPS techniques allows for quantum discord between system and environment. Hence, even if an initial state contains quantum discord, one can still obtain a homogeneous master equation in ECPS techniques. Finally, in the framework of homogeneous master equation, we show that the ECPS  techniques can yield convergence results for more general initial states. 

Since the relevant part in the new approach allows for quantum discord between system and environment, it may be helpful to consider the impact of  quantum discord. In fact, the new relevant part can contain all the correlations except for entanglement. Since entanglement is monogamy and the separable state is typical. The ECPS techniques  should have wide range of application due to quantum  de Finetti's theorem. Moreover,  even for the initial states that contain nontrivial entanglement between system and environment. One may still be able to restrict the inhomogeneous  term with monogamy properties of entanglement.  The   inhomogeneous  terms of ECPS techniques deserve further study. 

In this paper, we only show that the best projection superoperator can be dependent on the initial state. But the exact relationship is still absent. This issue needs further research. One may solve it by exploring the null space of $\mathcal{P}_\theta\mathcal{K}^2_{\text{tot}}-\mathcal{K}_2(\theta)$.

\begin{acknowledgments}
Financial support from National Natural Science Foundation of China under Grant
Nos. 11725524, 61471356 and 11674089 is gratefully acknowledged.
\end{acknowledgments}

\appendix
\begin{widetext}
\section{Simplification of the generator}\label{STG}
The term $ \langle\mathcal{P}_\theta\mathcal{L}_1(t)\mathcal{L}_1(t_1)\mathcal{P}_\theta \rangle$  in $\mathcal{K}^i_{1}(t)$ gives
\begin{align}\label{STDIG0}
  \langle\mathcal{P}_\theta\mathcal{L}_1(t)\mathcal{L}_1(t_1)\mathcal{P}_\theta\mathcal{O}\rangle= (1-\xi)^2\langle-  \mathcal{P}_\theta \sigma_+[B(t)(\mathcal{P}_\theta\mathcal{O})B^\dagger(t_1)]\sigma_-+ \mathcal{P}_\theta \sigma_+ [B(t)B^\dagger(t_1)(\mathcal{P}_\theta\mathcal{O})]\sigma_- \notag\\
  -  \mathcal{P}_\theta \sigma_-[B^\dagger(t)(\mathcal{P}_\theta\mathcal{O})B(t_1)]\sigma_++ \mathcal{P}_\theta \sigma_-[B^\dagger(t)B(t_1)(\mathcal{P}_\theta\mathcal{O})]\sigma_++\Hc \rangle.
\end{align}
It's easy to verify that 
\begin{align}\label{STDIG1}
  \mathcal{P}_\theta B(t)(\mathcal{P}_\theta\mathcal{O})B^\dagger(t_1)=\frac{1}{N}\mathcal{P}_\theta \Tr(\Pi^1B(t)(\mathcal{P}_\theta\mathcal{O})B^\dagger(t_1))\otimes\Pi^1\notag\\
  =\frac{1}{N^2}\Tr(\Pi^1B(t)\Pi^2B^\dagger(t_1))\mathcal{P}_\theta \Tr(\Pi^2\mathcal{P}_\theta\mathcal{O})\otimes\Pi^1=\frac{1}{N^2}\Tr(\Pi^1B(t)\Pi^2B^\dagger(t_1))\langle\mathcal{P}_\theta B(0)(\mathcal{P}_\theta\mathcal{O})B^\dagger(0)\rangle
\end{align}
 and
 \begin{equation}\label{STDIG2}
     \mathcal{P}_\theta B(t)B^\dagger(t_1)(\mathcal{P}_\theta\mathcal{O})=\frac{1}{N}\Tr(\Pi^1B(t)B^\dagger(t_1))\mathcal{P}_\theta \Pi^1(\mathcal{P}_\theta\mathcal{O})  =\frac{1}{N^2}\Tr(\Pi^1B(t)B^\dagger(t_1))\langle\mathcal{P}_\theta B(0)B^\dagger(0)(\mathcal{P}_\theta\mathcal{O})\rangle.
 \end{equation}
 Similar,  it's easy to find that
 \begin{equation}\label{STDIG3}
  \begin{aligned}
  \mathcal{P}_\theta B'(t)(\mathcal{P}_\theta\mathcal{O})B'^\dagger(t_1)=\frac{1}{N^2}\Tr(\Pi^+B'(t)\Pi^-B'^\dagger(t_1))\langle\mathcal{P}_\theta B'(0)(\mathcal{P}_\theta\mathcal{O})B'^\dagger(0)\rangle, \\
  \mathcal{P}_\theta B'(t)B'^\dagger(t_1)(\mathcal{P}_\theta\mathcal{O})=\frac{1}{N^2}\Tr(\Pi^+B'(t)B'^\dagger(t_1))\langle\mathcal{P}_\theta B'(0)B'^\dagger(0)(\mathcal{P}_\theta\mathcal{O})\rangle.
 \end{aligned} 
 \end{equation}
Those  two-point environmental correlation function  satisfy
\begin{align}\label{TPECF}
  \langle \Tr(\Pi^1B(t)\Pi^2B^\dagger(t_1)) \rangle=\langle \Tr(\Pi^+B'(t)\Pi^-B'^\dagger(t_1)) \rangle=\langle \Tr(\Pi^1B(t)B^\dagger(t_1))\rangle \notag \\
=\langle \Tr(\Pi^+B'(t)B'^\dagger(t_1))\rangle=\sum_{n_1,n_2}\exp(-i\omega(n_1,n_2)\tau)=\gamma N^2 h(\tau),
\end{align}
where $\tau=t-t_1$, $\gamma=2\pi /\delta\epsilon$ and 
  $h(\tau)=\sin^2(\delta\epsilon\tau/2)/(\pi\delta\epsilon\tau^2)$.
  Combining  \cref{TGAGT,STDIG0,STDIG1,STDIG2,STDIG3,TPECF}, we obtain
   \begin{equation}\label{SOTCLE}
    \frac{d}{dt}\mathcal{P}_\theta\rho_{i}(t)=\mathcal{K}_2(\theta)\rho_{i}(t)=\alpha^2\gamma \langle\mathcal{P}_\theta\mathcal{L}\mathcal{L}\mathcal{P}_\theta\rangle\mathcal{P}_\theta\rho_{i}(t) 
    =\alpha^2\gamma \mathcal{P}_\theta(\langle\mathcal{L}_1\mathcal{L}_1\rangle+\langle\mathcal{L}_2\mathcal{L}_2\rangle)\mathcal{P}_\theta\rho_{i}(t).
   \end{equation}

   The term $\mathcal{P}_\theta\langle\mathcal{L}_1(t)\mathcal{L}_1(t_1)\rangle$  in $ \mathcal{P}_\theta\mathcal{K}^2_{\text{tot}}$ gives
   \begin{align}\label{STDIG5}
    \mathcal{P}_\theta\langle\mathcal{L}_1(t)\mathcal{L}_1(t_1)\mathcal\rangle\mathcal{O}= (1-\xi)^2\langle-  \mathcal{P}_\theta \sigma_+[B(t)\mathcal{O}B^\dagger(t_1)]\sigma_-+ \mathcal{P}_\theta \sigma_+ [B(t)B^\dagger(t_1)\mathcal{O}]\sigma_- \notag\\
    -  \mathcal{P}_\theta \sigma_-[B^\dagger(t)\mathcal{O}B(t_1)]\sigma_++ \mathcal{P}_\theta \sigma_-[B^\dagger(t)B(t_1)\mathcal{O}]\sigma_++\Hc \rangle.
  \end{align}
It's easy to  verify that 
\begin{equation}
  \begin{aligned}
    \mathcal{P}_\theta\int_0^{t}dt_2B(t)\rho B^\dagger(t_2)=\frac{1}{N}\mathcal{P}_\theta\int_0^{t}dt_2\Tr(\Pi^1B(t)\rho B^\dagger(t_2))\otimes\Pi^1,\\
\mathcal{P}_\theta\int_0^{t}dt_2B'(t)\rho B'^\dagger(t_2)=\frac{1}{N}\mathcal{P}_\theta\int_0^{t}dt_2\Tr(\Pi^+B'(t)\rho B'^\dagger(t_2))\otimes\Pi^+.
  \end{aligned}
\end{equation}
  And 
  \begin{equation}
    \begin{aligned}
      \Tr(\Pi^1B(t)\rho B^\dagger(t_2))=\sum_{n_1,n_2}\exp(-i\omega(n_1,n_2)\tau)\langle n_2,2|\rho|n_2,2\rangle,\\
  \Tr(\Pi^+B'(t)\rho B'^\dagger(t_2))=\sum_{n_1,n_2}\exp(-i\omega(n_1,n_2)\tau)\langle n_2,-|\rho|n_2,-\rangle.
    \end{aligned}
  \end{equation}
  when $\tau\delta\epsilon\gg1$, the  time-dependent part  may be approximated as 
  \begin{equation}
    \sum_{n_1}\exp(-i\omega(n_1,n_2)\tau)\sim N\frac{2\sin(\tau \delta\epsilon/2)e^{-i\tau \delta\epsilon(1/2-n_2/N)}}{\tau \delta\epsilon}\sim \gamma N\delta(\tau).
  \end{equation}
  Using this approximation, we have
  \begin{equation}\label{STDIG4}
    \begin{aligned}
      B(t)B^\dagger(t_2)=\sum_{n_1,n_2}\exp(-i\omega(n_1,n_2)\tau)|n_1,1\rangle \langle n_1,1|\sim\gamma N\delta(\tau)\Pi^1,\\
  B'(t)B'^\dagger(t_2)=\sum_{n_1,n_2}\exp(-i\omega(n_1,n_2)\tau)|n_1,+\rangle \langle n_1,+|\sim\gamma N\delta(\tau)\Pi^+.
    \end{aligned}
  \end{equation}
  Combining  \cref{TGAGTT,STDIG4,STDIG5}, we obtain
   \begin{equation}
    \mathcal{P}_\theta\mathcal{K}^2_{\text{tot}}(t)\rho_{i}(t)=\alpha^2\gamma \mathcal{P}_\theta \langle\mathcal{L}\mathcal{L}\rangle\rho_{i}(t) 
    =\alpha^2\gamma \mathcal{P}_\theta(\langle\mathcal{L}_1\mathcal{L}_1\rangle+\langle\mathcal{L}_2\mathcal{L}_2\rangle)\rho_{i}(t).
   \end{equation}

  \section{The dynamic equation}\label{TDE}
 If set $\theta=0$, \cref{SOTCLE} gives
 \begin{equation}\label{P01DE}
  \begin{aligned}
    \frac{d}{dt}(\rho_{i}^{10,10}+\rho_{i}^{20,20}+\rho_{i}^{11,11}+\rho_{i}^{21,21})=0,\\
    \frac{d}{dt}(\rho_{i}^{21,21}+\rho_{i}^{10,10})=0,\quad \frac{d}{dt}(\rho_{i}^{21,21}-\rho_{i}^{10,10})=-\lambda\frac{\xi^2}{2}(\rho_{i}^{21,21}-\rho_{i}^{10,10}),\\
    \frac{d}{dt}(\rho_{i}^{20,20}-\rho_{i}^{11,11})=\lambda(-2+4\xi-\frac{5\xi^2}{2})(\rho_{i}^{20,20}-\rho_{i}^{11,11}),\\
    \frac{d}{dt}(\rho_{i}^{20,21}-\rho_{i}^{10,11})=\lambda(-\frac{1}{2}+\xi-\xi^2)(\rho_{i}^{20,21}-\rho_{i}^{10,11}),\\
    \frac{d}{dt}(\rho_{i}^{20,21}+\rho_{i}^{10,11})=\lambda(-\frac{1}{2}+\xi-\frac{3\xi^2}{2})(\rho_{i}^{20,21}+\rho_{i}^{10,11}),
   \end{aligned}
 \end{equation}
 where $\lambda=\alpha^2\gamma N$.

 If set $\theta=\pi/4$, \cref{SOTCLE} gives
   \begin{equation}\label{P+-DE}
  \begin{aligned}
    \frac{d}{dt}(\rho_{i}^{10,10}+\rho_{i}^{20,20}+\rho_{i}^{11,11}+\rho_{i}^{21,21})=0,\quad \frac{d}{dt}\rho_{i}^{10,21}=0,\\
    \quad\frac{d}{dt}(\rho_{i}^{10,20}+\rho_{i}^{10,11}+\rho_{i}^{20,21}+\rho_{i}^{11,21})=-\lambda\frac{1}{2} (\xi -1)^2(\rho_{i}^{10,20}+\rho_{i}^{10,11}+\rho_{i}^{20,21}+\rho_{i}^{11,21}),\\
    \frac{d}{dt}(\rho_{i}^{10,20}-\rho_{i}^{10,11}-\rho_{i}^{20,21}+\rho_{i}^{11,21})=\lambda(-\frac{5 \xi ^2}{2}+\xi -\frac{1}{2})(\rho_{i}^{10,20}-\rho_{i}^{10,11}-\rho_{i}^{20,21}+\rho_{i}^{11,21}),\\
    \frac{d}{dt}(\rho_{i}^{10,10}+\rho_{i}^{20,20}-\rho_{i}^{11,11}-\rho_{i}^{21,21})=\lambda(-\frac{3 \xi ^2}{2}+2 \xi -1)(\rho_{i}^{10,10}+\rho_{i}^{20,20}-\rho_{i}^{11,11}-\rho_{i}^{21,21}),\\
    \frac{d}{dt}(\rho_{i}^{10,20}-\rho_{i}^{11,21})=\lambda(-\xi ^2+\xi-\frac{1}{2})(\rho_{i}^{10,20}-\rho_{i}^{11,21}).
  \end{aligned}
\end{equation}
\end{widetext}

\end{document}